\documentstyle[12pt,aasms4]{article}
\begin{document}

\title{Disk Formation In 
Hierarchical Hydrodynamical Simulations: A Way Out Of The Angular
 Momentum Catastrophe}

\author{ R. Dom\'{\i}nguez-Tenreiro, P.B. Tissera\altaffilmark{1} and A. 
S\'aiz} 
\affil{Dpt. F\'{\i}sica Te\'orica C-XI, 
Universidad Aut\'onoma de Madrid, \\
E-28049 Cantoblanco, 
Madrid, 
Spain;\\ 
rosa@astrohp.ft.uam.es, patricia@iafe.uba.ar, asaiz@delta.ft.uam.es}
\altaffiltext{1}{Visitor at Imperial College of Science, Technology and
Medicine, London, UK\\ Present address: IAFE, Casilla de Correos 67, Suc. 28,
1428 Buenos Aires, Argentina} 

\begin{abstract}
We report results on the formation of disk-like structures in
two cosmological hydrodynamical
simulations in a hierarchical clustering scenario, sharing
the same initial conditions. In the first one, 
a simple and generic implementation of star formation 
 has allowed galaxy-like objects with stellar bulges and extended,
populated disks to form. Gas in the disk comes from both, particles
that survive mergers keeping in part their angular momentum content, and
new gas supply by infall, once the merger process is over, with
global specific angular momentum
conservation. The stellar bulge forms from gas that has lost most of its
angular momentum.
In the second simulation, no star formation has been included.
In this case, objects consist of an overpopulated central gas concentration,
and an extended, underpopulated disk.
The central concentration forms from particles that
suffer an important angular momentum loss in violent events,
and it often contains more than 70$\%$ of the object's baryonic mass.
The external disk forms by late infall of gas, that roughly
conserves its specific angular 
momentum. The difference between these two simulations
is likely to be due to the stabilizing character of the stellar
bulge-like cores that form in the first simulation, which diminishes the 
inflow of gas triggered by mergers and interactions.

\end{abstract}


\section{INTRODUCTION}
The standard model of disk formation has been set by Fall and Efstathiou
(1980) following and extending previous work by White and
Rees (1978). According to this scheme, extended disks resembling those
observed in spiral galaxies can be formed from the
 diffuse halo gas component provided that it conserves its specific angular
momentum ($j$) during collapse.
 This scheme has served as a basis to build-up 
theoretical models of galaxy formation (Lacey et al. 1993; 
Kauffmann 1996; Dalcanton et al. 1997; Mo et al. 1998; van den Bosch 1998,
hereafter vdB98),
but it has the shortcoming of not being able to treat the effects of mergers
in disk formation. Merger effects are naturally taken into account 
in numerical simulations.
However, so far, no hydrodynamical
simulation of galaxy formation in fully consistent hierarchical 
cosmological scenarios had been able
to produce extended disks, with structural and dynamical properties 
similar to those of observed spirals.
The problem was the excessive loss of angular momentum by the gas clumps 
as they merge inside the dark haloes, resulting in too concentrated disks 
 (the so-called angular momentum catastrophe problem, hereafter AMC, see
 Navarro \& Benz 1991; Evrard et al. 1994; Vedel et al. 1994; 
 Navarro et al. 1995; Navarro \&
Steinmetz 1997; Weil et al. 1998). 
No star formation processes have been considered by these 
authors. By contrast, the 
effects of star formation
have been considered by Katz (1992) and 
Steinmetz \& M$\ddot{\rm u}$ller (1995)
in a semi-cosmological modellization
of the collapse of an {\it isolated} constant density perturbation in
{\it solid-body} rotation,  
getting in both cases  a three component
system that resembles a spiral galaxy. 

A realistic implementation of the star-forming processes in a simulation is 
beyond present possibilities. However, turning on 
the star formation process could have 
important 
consequences in building up galaxy-like objects in hierarchical scenarios,
since stellar bulge-like cores would be formed
at high redshifts, modifying substantially
the fate of the subsequently formed disk-like objects
relative to 
that of their bulgeless counterparts
(Mihos \& Hernquist 1994, 1996, hereafter MH94, MH96).
In fact,
both theoretical studies and numerical simulations have shown that a disk
can develop violent instabilities leading to bar formation, followed by
inward material transport due to $j$ non conservation
(Toomre \& Toomre 1972; Ostriker \& Peebles 1974; Toomre 1981;
Efstathiou et al. 1982; Athanassoula \& Sellwood 1986; Binney \&
Tremaine 1987; 
Barnes \& Hernquist 1991, 1992;
Friedli \& Benz 1992;
Martinet 1995 and references 
quoted
therein;
 MH94, MH96). However, Athanassoula \& Sellwood 1986, MH94, MH96, 
 Christodoulou et al. (1995) and vdB98,
have shown that bulges, if present, play a
fundamental role in stabilizing disk galaxies against the bar instability
mode, diminishing the inflow of gas triggered by 
interactions and mergers.

Hence, stellar bulges could be critical to ensure global 
$j$ conservation in the assembly of disks in hydrodynamical
cosmological simulations.
The inclusion of a star formation algorithm leads unavoidably
to stellar bulge formation, regardless of its details.
So even a simple implementation, that takes into account only its gross
physical properties, provided that not all the gas is depleted
at high redshifts, could be enough to ensure the formation of disks
and their stability.
In this Letter we discuss the problem of the $j$ conservation 
in connection with disk formation in the hierarchical clustering scenario.
We analyze the role played by central compact bulges as a critical component
to prevent the AMC.
\section{DISK FORMATION} 
We have followed the evolution of $64^3$ particles in a periodic box of
10 Mpc ($H_0 = $50 km s$^{-1}$ Mpc$^{-1}$) using a SPH code coupled to the 
high resolution AP3M code (Thomas \& Couchman 1992), either 
including a star formation algorithm (S1 simulation) or not 
(S2 simulation). 
The initial distribution of positions and velocities is the same
in both S1 and S2, and is consistent with 
a standard flat CDM cosmology, with $ \Omega_{\rm b} = 0.1,
\Lambda = 0 $ and $b = 2.5$. All, dark, gas and star particles have the same
mass, $m = 2.6 \times 10^8$ M$_{\odot}$. The integrations were carried out 
only gravitationally from $z \simeq 50$
to $z = 10$, and from there on the SPH forces were also taken into account
up to $z = 0$
using fixed time steps of $\Delta t = 1.3 \times 10^7$ years.
The gravitational  softening length is 3 kpc and the minimum allowed 
smoothing length is 1.5 kpc.
In S1, cold and dense 
($\rho_{\rm gas} > 7 \times 10^{-26}$ g/cm$^{3}$)
gas particles which also satisfy the Jeans
instability criterion are transformed into stars according to:
$d {\rho}_{\rm star}/ dt = - c \rho_{\rm gas}/t_{\star}$, 
with star formation efficiency $c = 0.01$ and $t_{\star}$ a 
characteristic time scale (for  details,
see Tissera et al. 1997).
No supernovae explosion effects have been considered.
The low $c$ value used implies that star formation occurs mainly in the
very dense regions, and so it allowed us to 
have available gas to form disk-like structures at low $z$. 
To some extent, 
this could 
mimic the effects of energy injection from supernovae explosions.
Haloes formed in S1 are identical to those formed in S2 
 (Tissera \&
 Dom\'{\i}nguez-Tenreiro 1998, hereafter TDT98). The
baryonic disk-like objects (DLOs) that they host have been 
identified using a friend-of-friends algorithm. 
 Only those DLOs whose total baryon number
(i.e., star, $N_{\rm star}$, plus gas, $N_{\rm gas}$, number) satisfies
$N_{\rm baryon} > 150$ have been considered. The number of particles per
DLO is essentially the same in the S1 and S2 simulations.
Ten disks (and three spheroids, not considered in this paper)
have been found in S1 following this 
criterion.
Their general characteristics are given in Table 1. DLOs \#5 and \#6 form
a pair in S1, while in S2 they cannot be resolved as two distinct objects.

DLOs that form in S1 have central star concentrations and extended,
populated disks, while those in S2 present an inner, rather disordered gas 
concentration and, also, extended disks, but with a much lower 
surface mass density than their S1 counterparts. The mass density profiles of baryons,
projected on the disk plane, are well fit in both S1 and S2 by a double
exponential (Courteau et al. 1996; Courteau 1997) whose bulge 
and disk scale lengths, $R_{\rm b}$ and
$R_{\rm d}$, respectively, are given in Table 1. As a measure of the
central baryonic concentrations, in Table 1 we also give the corresponding
{\it mass} $B/D$ ratios. These are larger for S2 versions of the DLOs,
although the differences vary from DLO to DLO.

In Fig. 1 we plot the specific 
total angular momentum at $z = 0$ versus mass for
dark haloes in S1 or S2, $j_{\rm dh}$,
for the inner 83\% of the gas mass for S1 and S2 DLOs
(i.e., the mass fraction
enclosed by $R_{\rm opt} = 3.2R_{\rm d}$ in a purely exponential disk),
$j_{\rm g}$, and for the 
stellar component in S1, $j_{\rm s}$. We see that
 $j_{\rm g}$ is of the order of 
 $j_{\rm dh}$ for S1 DLOs, so that these gas particles have collapsed
conserving, on average, their angular momentum,
while those in their S2 counterparts have suffered a strong
loss, in agreement with previous results. 
Moreover, DLOs formed in S1 are inside the box defined by observed spiral 
disks in this plot (Fall 1983), while DLOs in S2 are not. The stellar 
component in S1 has formed from gas that had lost a
substantial fraction of its $j$.
To make these points clearer,
in Fig. 2a we plot, for each baryon particle of halo hosting 
DLO \#1 in S1, its angular 
momentum component per unit mass $j_{z,i}$ 
(parallel to the total angular momentum of the 
disk, $\vec{J}_{\rm dis}$) versus $R_i$, the projected 
radial distance from $i$ to the mass center, 
at $z = 0$. The full line is $v_{\rm c}(R)R$, where $v_{\rm c}(R)$
is the circular 
velocity at $R$ in the potential well of both the dark matter and baryons 
(see TDT98). We see that most gas particles 
placed at $R_i \stackrel{<}{\scriptstyle \sim} 30$ kpc follow circular 
trajectories on the equatorial 
plane, that is, they have
$j_{z,i} \simeq |\vec j_i| \simeq  v_{\rm c}(R_i)R_i$,
with a small 
dispersion around this value, so that they form a cold thin disk. In contrast, 
those at $R_i \stackrel{>}{\scriptstyle \sim} 30$ kpc (hereafter, halo gas 
particles) are disordered, with 
their $|j_{z,i}|$ taking any value under the full line. Roughly half halo gas 
particles have $j_{z,i} < 0$ (counterrotating particles, open circles).
Stars at $R_i \stackrel{<}{\scriptstyle \sim} 2$ kpc form a compact central 
relaxed core, with $\vec{j}_i$ 
without any preferred direction and very low $|\vec j_i|$, while those at $R_i 
\stackrel{>}{\scriptstyle \sim} 2$ kpc roughly follow a (thicker) disk.
These same
plots for DLO \#1 in S2 (Fig. 2b), show that their gas particles 
placed at 6 kpc $\stackrel{<}{\scriptstyle \sim} R_i \stackrel{<}{\scriptstyle 
\sim}$ 30 kpc form, also in this case, an extended, thin, 
ordered disk, but now most gas particles are at $R_i \stackrel{<}{\scriptstyle 
\sim} 6$ kpc and show an 
important dispersion around small $j_{z,i}$ values, even with negative ones. 
Halo gas particles show similar properties in both simulations.
The behaviour patterns for baryons at $z = 0$ described so far are common to 
the other DLOs identified either in S1 or S2. In any case, gas particles in 
the cold disk component have 
globally conserved their $j$,
while most of those in the central regions (in S2) 
or those giving rise to stars in the bulge (in S1) have been 
involved in an inflow event with high $j$ loss, i.e., in an AMC.

The first of these two processes can be understood as follows: the net effect 
of shocks and cooling on disordered halo gas particles in the stages of 
quiescent evolution is that the fluid is forced to a coherent rotation with 
global specific angular momentum conservation. Consequently, 
gas particles tend to settle at the halo center, moving on a plane (the equatorial 
plane) on circular orbits if the gravitational potential at the inner regions 
is axisymmetric. 
The axisymmetric character of the potential 
is self-regulatory, as in the inner regions baryons are dynamically dominant 
(see TDT98). So, cold thin disks naturally 
appear in the non-violent phases of evolution, as in Fall \& Efstathiou (1980).

However, as previously stated, 
cold disks are known to be strongly unstable against the bar  
instability mode. Massive dark haloes can stabilize disks, 
but some analytical works on disk stability (Christodoulou et al. 1995; vdB98) 
 suggest that
not every halo is able to stabilize any 
amount of baryons as a pure exponential disk, and a bulge is 
needed to ensure stability. In the absence of a bulge, the disk would develop a 
bar instability, 
that implies a loss of the axial character of the gravitational 
field produced by the baryon component and so, $j_{z,i}$ is no more conserved 
and the gas particles can fall to the center. This could be the process at work 
in the angular momentum losses observed in S2 (and the star forming gas in S1) 
and other author's simulations.
Global disk stability is usually studied through the $X_2(R)$ parameter (Toomre 1981,
Binney \& Tremaine 1987). We 
have calculated $X_2(R)$ for the disk component of our DLOs at different $z$. 
Moreover, to find out whether haloes formed in S1 or S2 need a bulge to stabilize pure 
exponential disks with baryon masses as deduced from Table 1, we have also 
calculated $X_2(R)$ for bulgeless versions of our DLOs, i.e., putting all the 
baryonic mass of each DLO in a pure exponential disk, whose
 scalelength is determined assuming specific angular momentum conservation. 
In Fig. 2a (2b) we 
plot $X_2(R)$ for the S1 (S2) version of DLO \#1 at $z = 0$.
 The same plots for the same DLO are given after 
its last major event in Fig. 2c (2d) for its S1 (S2) version,
and previous to this event in Figs. 2e (for S1) and 2f (for S2).
Moreover, the
$X_2(R)$ for their bulgeless counterpart at $z = 0$ is plotted
 in Fig. 2b.
Recalling the $X_2(R)$ stability criterion,
if we define $R_{\rm st}^{\rm ad}$ and $R_{\rm st}^{\rm ped}$
as the points where $X_2(R)=3$ (stability thresholds) 
for actual and pure exponential disks,
respectively, it is apparent from these Figures that disks, when present, are
stable: they are
detected at $R > R_{\rm st}^{\rm ad}$
if they have had enough time to form after the last merger
(see below). By contrast, the bulgeless version of DLO \#1 
at $z = 0$ would be stable only at larger $R$ ($R >
R_{\rm st}^{\rm ped} \simeq 21$ kpc).
 This behaviour is 
common to any DLO in S1 or S2, and so central mass concentrations are needed
to stabilize these disks. 

The role played by stellar cores will be better clarified through a description 
of how galactic-like objects are built-up in our simulations. Their assembly in 
S1 is an inside-out process with different episodes.
i) First, dark matter haloes collapse at high $z$ 
forming a first generation of (small) disks and stars.
ii) Then, the first unstabilizing mergers at high $z$ happen,
resulting in disk disruption and rapid 
mass inflow to the central regions with $j$ loss and violent star 
formation, mainly at the central regions.
Also, most preexisting stars will concentrate at the 
center of the new object through violent relaxation. These two processes help 
build up a central stellar bulge-like structure.
iii) After the first mergers, a 
disk is regenerated through an infall of gas particles, either belonging to the 
baryonic merging clumps or diffuse, as previously described. For example, a 
compact stellar bulge and an almost cold disk in S1 DLO \#1 at $z = 0.57$ are 
apparent in Fig. 2e.
iv) After disk regeneration, the system can undergo new major 
merger events at lower $z$
(see Tissera et al. 1998 for details).
During the orbital decay phase, previous to the actual fusion of the DLOs, most 
of their orbital angular momentum 
is transported to (the particle components of) 
each host halo, spinning it up (Barnes 1992; Barnes \& Hernquist 1996, hereafter
BH96).
Because,
now, the disks involved in 
the merger are stabilized by their bulges, no strong gas inflow occurs in this 
phase (MH96). As the disks approach one another, 
they are heated and finally disrupted, but the high 
efficiency of gas shocking and cooling, and the symmetry of the central potential,
quickly puts those of their gas particles 
with high $J_i$ into a new intermediate disk, 
while their low $J_i$ particles sink to the center
where most of them are transformed into 
stars, feeding the bulge. The stellar bulge of the smaller DLO is eventually 
destroyed and incomplete orbital angular momentum loss puts 
most of its stars on the 
remnant disk (Fig. 2c, note incomplete relaxation).
v) Relaxation and disk regeneration are completed. Most of disk external 
particles are supplied by infall, as in iii) (Fig. 2a).

The assembly of galactic-like objects in S2 follows the same stages. We recall 
that in both simulations, haloes and merger trees are identical. The main 
difference is that in S2, the i) and ii) stages do not result 
in a stellar core, 
and, consequently, in iii) stage an unstable gas disk is formed, susceptible 
to grow bars. In particular, during the orbital decay phase in iv),
 strong gas inflow and $j$ loss are induced (Fig. 2f, see also MH96 and
BH96). 
The actual fusion completes the gas inflow (Fig. 2d),
 involving most of the gas particles originally in the disks.
Few of them are left for disk regeneration, so that, in phase v), disks 
are formed almost only from halo gas particles 
(Fig. 2b). Small satellites that orbit around DLOs may also trigger,
in this case, 
a further gas inflow (as in MH94), 
while in S1 they are accreted without any major damage.
We recall that for a given choice of the star formation 
parameters, the final characteristics of one object in a simulation 
are determined by its merger tree as well as the particular 
values that the parameters 
describing mergers and interactions take in each event (Barnes 1992, MH94,
BH96, MH96).
This explains the dispersion in the $B/D$ values among the different 
DLOs in S1 (or S2).

 Concerning numerical resolution, DLOs in S1 and S2
 are resolved with a relatively low number of particles. In contrast,
dark matter haloes are described with a much better resolution.
An inappropiate low gas resolution would result in an unphysical gas
heating that could halt the gas collapse (Navarro \& Steinmetz 1996).
However, some works suggest that it is an inadequate resolution in the dark
matter halo component that may produce the larger undesired numerical
artifacts (Steinmetz \& White 1997). Therefore, to make sure that
the populated and extended disks in S1 do not
result from unphysical gas heating, we have
run a higher resolution simulation ($64^{3}$ particles 
in a periodic box of 5 Mpc, with 
cosmological and star formation parameters similar
to those in S1; hereafter HRS). Only one disk with mass comparable to those
in S1 forms. Its analysis has shown that it is populated and extended,
that its 
structural and dynamical characteristics are compatible 
with observations (see Table 1 and
Fig. 1) and that the physical processes leading to its formation
are essentially the same as those that are at work in S1.
In addition, a comparison between the distributions of the ratios
$t_{\rm dyn}/t_{\rm cool}$ for gas particles belonging to the 
DLOs in S1 and to their counterparts in S2 shows no difference. These results
indicate that the infall of gas in S1 has not been artificially affected 
 by the decrease of numerical resolution due to the transformation of gas 
particles into stars.

To summarize, a simple implementation of star formation that prevents 
 gas depletion at high redshifts, but permits the formation of stellar bulges,
has allowed extended and populated disks to form at later 
times. 
In a more realistic model, supernovae should play this part, leading to a 
self-regulating star formation.
These disks have masses and specific angular momenta compatible with observed
spirals, and their bulge and disk scales are also consistent with their
observable values (Courteau, de Jong \& Broeils 1996; Courteau 1997).
On the contrary, if the implementation of star formation had resulted
in an early gas depletion into stars as it cools and collapses,
no gas would have been left to form new disks 
(TDT98; Steinmetz \& Navarro 1998). The generality and
simplicity of the implementation we have used
suggests that extended, populated disks are generic
(Silk \& Wyse 1993), and that they 
easily form when $j$ loss
in gas collapse and mergers is prevented by stabilizing the disks with
a stellar bulge.

This work was partially supported by the
DGICyT, Spain (grants number PB93-0252, PB96-0029), which also
supported P. Tissera and A. S\'aiz through fellowships.
P. Tissera would thank the Astrophysic Group at 
ICSTM (London) for their hospitality.
We are indebted to the Centro de Computaci\'on Cient\'{\i}fica (Universidad
Aut\'onoma de Madrid) and to the Oxford University 
for providing the computational support to perform 
this paper.

\clearpage

\begin{deluxetable}{cccccc}
\tablecaption{SOME CHARACTERISTICS OF DLOS\tablenotemark{a}}
\tablehead{
\colhead{DLO}   & \colhead{ $N_{\rm gas}$}  &
\colhead{$N_{\rm star}$}  &
\colhead{$B/D$}    & \colhead{ $R_{\rm b}$} &
\colhead{$R_{\rm d}$}}
\startdata
1& 348  &278  &1.19(3.66)&0.74(1.29)&7.33(6.99)
\nl
2& 359 & 240  &1.19(2.17)&0.74(1.29)&5.66(7.08)
\nl
3& 307 & 211
&1.55(1.75)&0.85(1.41)&10.90(14.04)
\nl
4& 311 & 215  &1.60(4.20)&0.74(1.19)&9.98(9.02)
\nl
5& 210 & 95 &1.15      &0.54      &6.50
\nl
6& 151 & 69 &1.22      &0.53      &5.61
\nl
7& 227 & 79  &2.02(2.87)&0.99(1.27)&6.56(5.87)
\nl
8& 189 & 157 &1.31(5.43)&0.49(1.32)&5.29(5.62)
\nl
9& 108 &  99 &2.95(4.36)&0.54(1.47)&7.07(13.31)
\nl
10&109 & 47  &1.30(2.05)&0.40(1.23)&9.75(10.75)
\nl
HRS&1713 & 1380  &0.86      &1.13      &9.58
\nl

\enddata
\tablenotetext{a}{Distances are given in kpc;
quantities in parentheses correspond to S2 DLOs.
DLO masses are given by
$M_{\rm DLO} = m \times (N_{\rm gas} + N_{\rm star})$.}

\end{deluxetable}

\clearpage

\begin{figure}
\figcaption{The specific angular momentum at $z = 0$ versus the mass, for 
haloes in S1 
or S2 ({\it filled circles}), and HRS ({\it filled square}); 
the inner 83\% of the gas component
in S1 DLOs ({\it open triangles}), S2 DLOs ({\it filled triangles})
and the HRS disk ({\it open square}); and the stellar
component in S1 ({\it open stars}) and HRS ({\it asterisk}). 
  The {\it solid (dotted)} box shows the 
region occupied by
the spiral disks (ellipticals).} 
\label{fig1}
\end{figure}

\begin{figure}
\figcaption{
Specific angular momentum component along $\vec{J}_{\rm dis}$ for each baryon
particle of halo \#1, versus their positions at different $z $.
{\it Circles}: gas particles, {\it  stars}: stellar particles;
{\it open symbols}: counterrotating particles.
{\it Left panels}: S1 version at different $z$; {\it right panels}: S2 version at
approximately the same $z$.
{\it Full lines}: $v_{\rm c}(R) R$;
{\it dotted lines}: $X_2(R)$ for actual disks at each $z$;
{\it dashed line}: $X_2(R)$ for the pure exponential version at $z = 0$.
The {\it arrows} mark the point where $X_2(R) = 3$, i.e., 
$R_{\rm st}^{\rm ad}$ and $R_{\rm st}^{\rm ped}$.}
\label{fig2}
\end{figure}


\clearpage

\begin{thebibliography}{}

 
\bibitem{} 
Athanassoula, E., \& Sellwood, J. 1986, MNRAS, 221, 213 

\bibitem{}
Barnes, J.E., 1992, ApJ, 393, 484

\bibitem{}
Barnes, J.E., \& Hernquist, L. 1991, \apj, 370, L65; 1992, ARA\&A, 30, 705;
1996, \apj, 471, 115 (BH96)

\bibitem{}
Binney, J., \& Tremaine, S. 1987, {\it Galactic Dynamics}, 
(Princeton: Princeton Univ. Press) ch. 6

\bibitem{}
Christodoulou, D.M., Shlosman, I., \& Tohline, J.E. 1995, ApJ, 443, 551

\bibitem{}
Courteau, S. 1997, in {\it Morphology \& Dust Content in Spiral Galaxies}
eds. D. Block \& M. Greenberg, (Dordrecht: Kluwer) 

\bibitem{}
Courteau, S., de Jong, R.S., \& Broeils, A.H. 1996, \apj, 457, L73 

\bibitem{}
Dalcanton, J.J., Spergel, D.N., \& Summers, F.J. 1997, \apj, 482, 659

\bibitem{}
Fall, S.M. 1983, in IAU Symp. 100 {\it Internal Kinematics and Dynamics of 
Galaxies}, 
ed. E. Athanassoula (Dordrecht: Reidel), p. 391

\bibitem{}
Fall, S.M., \& Efstathiou, G. 1980, MNRAS, 193, 189 

\bibitem{}
Esfstathiou, G., Lake, G., \& Negroponte, J. 1982, MNRAS, 130, 125

\bibitem{}
Evrard, A.E., Summers, F.J., \& Davis, M. 1994, \apj, 422, 11

\bibitem{}
Friedli, D., \& Benz, W. 1993, \aap, 268,65

\bibitem{}
Katz, N. 1992, \apj, 391, 502

\bibitem{}
Kauffmann, G. 1996, MNRAS, 281, 475


\bibitem{}
Lacey, C., Guiderdoni, B., Rocca-Volmerage, B., \& Silk, J. 1993, \apj, 402, 
15


\bibitem{}
Martinet, L. 1995, Fund. Cosmic Phys., 15, 341


\bibitem{}
Mihos, J.C., \& Hernquist, L. 1994, \apj, 425, L13 (MH94); 1996, \apj, 464, 
641
(MH96)

\bibitem{}
Mo, H.J., Mao, S., \& White, S.D.M.  1998, MNRAS, 295, 319 

\bibitem{}
Navarro, J.F., \& Benz, W. 1991, \apj, 380, 320

\bibitem{}
Navarro, J.F., Frenk, C.S., \& White, S.D.M. 1995, MNRAS, 275, 56

\bibitem{}
Navarro, J.F., \& Steinmetz, M. 1997, \apj, 438, 13

\bibitem{}
Ostriker, J.P., \& Peebles, P.J.E. 1973, \apj, 186, 467


\bibitem{}
Silk, J., \& Wyse, R.M. 1993, Phys. Rep., 231, 293

\bibitem{}
Steinmetz, M., \& M$\ddot{\rm u}$ller 1995, MNRAS, 276, 549

\bibitem{}
Steinmetz, M., \& Navarro, J.F. 1998, SISSA astro-ph 9808076 preprint

\bibitem{}
Steinmetz, M., \& White, S.D.M. 1997, MNRAS, 289, 545

\bibitem{}
Thomas, P.A., \& Couchman, H.M.P. 1992, MNRAS, 257, 11


\bibitem{}
Tissera, P.B., \& Dom\'{\i}nguez-Tenreiro, R. 1998, MNRAS, 297, 177 (TDT98)

\bibitem{}
Tissera, P.B., Goldschmidt, P., \& Dom\'{\i}nguez-Tenreiro, R. 1998, MNRAS,
submitted

\bibitem{}
Tissera, P.B., Lambas, D.G., \& Abadi, M.G. 1997, MNRAS, 286, 384

\bibitem{}
Toomre, A., \& Toomre, J., 1972, \apj, 178, 623

\bibitem{}
Toomre, A. 1981, in {\it The Structure and Evolution of Normal Galaxies}, eds. 
S.M. Fall \& D. Lynden-Bell, (Cambridge: Cambridge Univ. Press), p. 111

\bibitem{}
van den  Bosch, F.C., 1998, SISSA astro-ph 980113 preprint (vdB98)

\bibitem{}
Vedel, H., Hellsten, U., \& Sommer-Larsen, J. 1994, MNRAS, 271, 743

\bibitem{}
Weil, M.L., Eke, V.R., \& Efstathiou, G., 1998, SISSA astro-ph 9802311 preprint
\bibitem{}
White, S.D.M., \& Rees, M.J. 1978, MNRAS, 183, 341



\end{thebibliography}
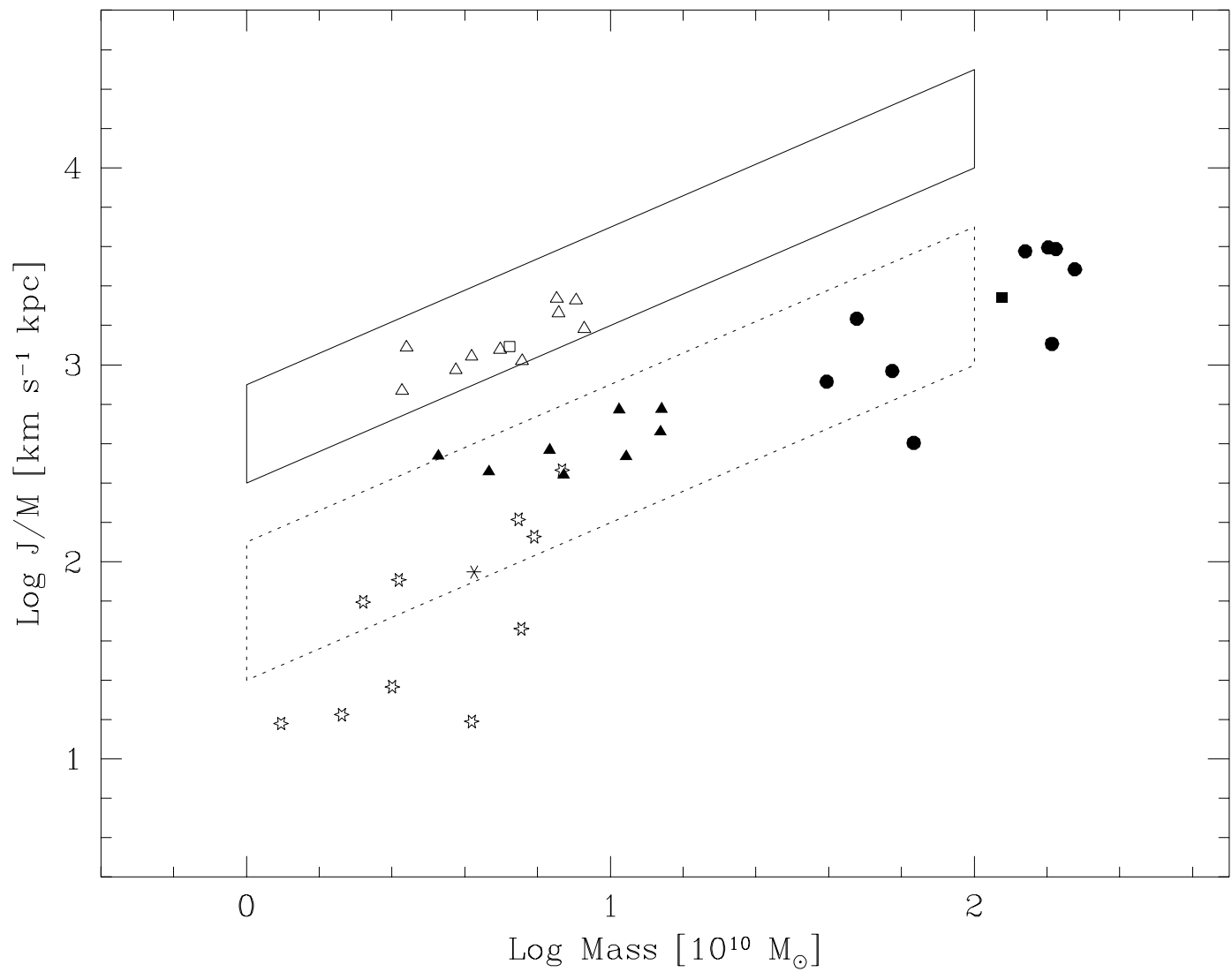
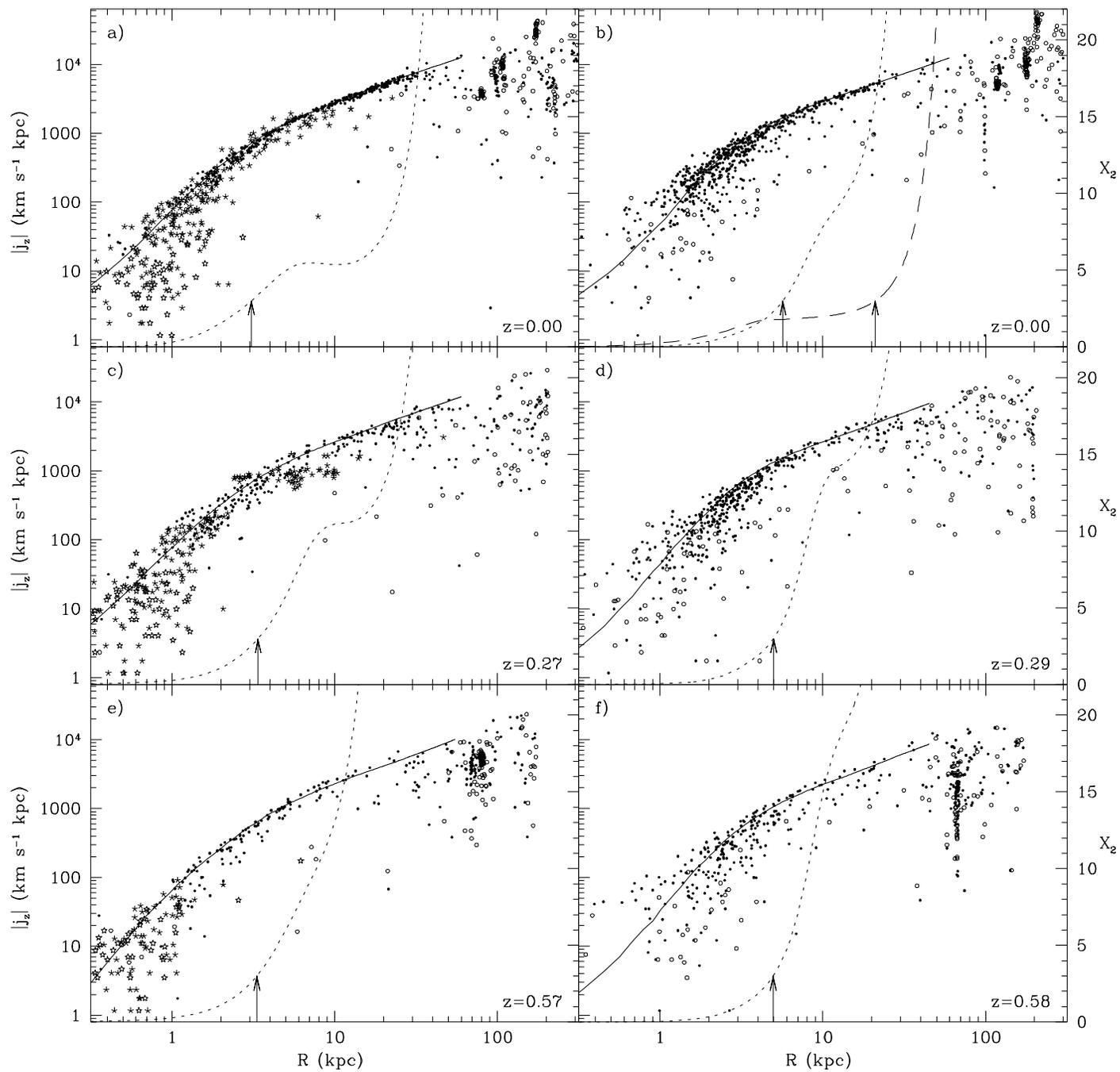
\end{document}